\documentclass[proof]{WileyASNA-v1}

\articletype{Original Article}

\newcommand{\mtot}{M_{\rm tot}}
\newcommand{\mdot}{\dot{M}}
\newcommand{\vcom}{\mathbf{v}_{\rm COM}}
\newcommand{\vtrans}{\mathbf{v}_{\rm trans}}
\newcommand{\rell}{\ell_1}
\newcommand{\dL}{\Delta L}

\usepackage{hyperref}

\begin{document}

\title{The Effective Velocity of Transferred Mass: How Momentum Prescriptions Determine Binary Orbital Evolution}

\author[1]{Jerry (Yuze) Li*}

\authormark{LI}

\address[1]{\orgname{Marc Garneau Collegiate Institute}, \orgaddress{\state{Toronto, ON}, \country{Canada}}}

\corres{*Jerry (Yuze) Li, Marc Garneau Collegiate Institute, Toronto, ON, Canada. \email{fireheartjerry@gmail.com}}

\abstract{In binary stellar evolution, the orbital response to mass transfer depends on how angular momentum is redistributed. We introduce a one-parameter family of prescriptions characterized by $\eta$, the fractional weight of the donor velocity in the effective velocity of the transferred mass: $\vtrans = \eta\,\mathbf{v}_1 + (1-\eta)\,\mathbf{v}_2$. We derive a closed-form expression for the angular momentum change per transfer event, $\dL/L = \delta m\,[(1-\eta)/M_2 - \eta/M_1]$. The two endpoint prescriptions ($\eta = 1$ and $\eta = 0$) produce angular momentum changes of opposite sign, yielding qualitatively different orbital evolution at every mass ratio. Conservative mass transfer ($\dL = 0$) corresponds uniquely to $\eta = M_1/\mtot$, i.e.\ $\vtrans = \vcom$. For constant~$\eta$, we derive the general closed-form solution $a_f/a_0 = [(1-f)^{2(1-\eta)}(1+fq_0)^{2\eta}]^{-1}$.}

\keywords{binaries: close -- methods: analytical -- methods: numerical -- stars: evolution -- stars: mass-loss}

\maketitle

\section{Introduction}\label{sec1}

The long-term orbital evolution of a mass-transferring binary is governed by the redistribution of angular momentum between the stellar components and any material leaving the system \citep[see][for comprehensive reviews]{Paczynski1971, Tauris2023, PostnovYungelson2014}. The standard treatment of angular momentum in mass-transferring binaries is developed in detail by \citet{FrankKingRaine2002}, who derive the orbital response under the assumption of conservative transfer, and by \citet{Soberman1997}, who extend the formalism to include non-conservative losses. Under the assumption that total orbital angular momentum is conserved during conservative (non-mass-losing) transfer, the classical result gives
\begin{equation}\label{eq:classical}
  \frac{\dot{a}}{a} = -2\,\frac{\mdot_1}{M_1}\!\left(1 - \frac{M_1}{M_2}\right),
\end{equation}
predicting orbital expansion when the donor is less massive than the accretor ($q < 1$) and contraction when $q > 1$ \citep{Soberman1997}. This result underpins the treatment of mass transfer in population synthesis codes such as BSE \citep{Hurley2002}, COSMIC \citep{Breivik2020}, and COMPAS \citep{TeamCOMPAS2022}.

It is important to distinguish the framework presented here from the non-conservative mass transfer formalism of \citet{Soberman1997}, who parameterize the angular momentum carried away by material \emph{lost from the system} through modes such as fast winds ($\alpha$), Jeans-mode ejection ($\beta$), and circumbinary ring outflow ($\delta$). Those parameters control the angular momentum budget of mass that \emph{leaves} the binary. The present work addresses a complementary question: how is momentum partitioned for mass that \emph{remains} in the system and is transferred from donor to accretor? We denote our prescription parameter $\eta$ (rather than $\alpha$) to avoid confusion with the mass-loss mode parameters of \citet{Soberman1997}. In principle, the two frameworks can be combined: the Soberman parameters govern non-conservative losses, while $\eta$ governs the momentum closure for the conservative component of the transfer. In short, \citet{Soberman1997} parameterize angular momentum loss when mass \emph{leaves} the system; $\eta$ parameterizes the momentum closure when mass \emph{remains} in the system.

Equation~(\ref{eq:classical}), however, encodes an implicit physical assumption: that angular momentum is \emph{exactly} conserved during each transfer event. This assumption is equivalent to requiring that the transferred mass carries the centre-of-mass velocity---a point noted implicitly in textbook derivations \citep{FrankKingRaine2002, Tauris2023} but not typically stated as an explicit physical condition. In practice, every numerical implementation must specify a concrete prescription for how momentum is partitioned when mass moves from donor to accretor. Different prescriptions correspond to different effective velocities carried by the transferred material, and therefore to different angular momentum budgets---even when the total mass is conserved.

The choice of prescription is not merely a technical detail. This degree of freedom is implicit in the standard derivation because the conservative assumption ($\dot{L}_{\rm orb} = 0$) is typically imposed before any parameterization is introduced; once $\dot{L}_{\rm orb} = 0$ is imposed, no free parameter remains and the underlying velocity condition is invisible. Because the angular momentum change determines whether an orbit expands or contracts, prescription-dependent differences can propagate into qualitatively different evolutionary outcomes: common-envelope inspiral versus stable Roche-lobe overflow \citep{Podsiadlowski1992, Ivanova2013}, formation versus disruption of compact-object binaries, and ultimately the predicted rate of gravitational-wave sources detectable by LIGO--Virgo--KAGRA \citep{Abbott2016, Mandel2022}. The efficiency of mass transfer---the fraction of transferred mass actually retained by the accretor---is itself a major uncertainty \citep{deMink2009}, and the orbital response depends on the combined effect of mass loss fraction \emph{and} the momentum prescription for the retained material. Quantifying the sensitivity of orbital evolution to the momentum closure is therefore a prerequisite for robust population synthesis.

In this paper we introduce a one-parameter family of momentum prescriptions, the $\eta$-family, that continuously interpolates between two natural endpoints: the donor-velocity prescription ($\eta = 1$) and the accretor-velocity prescription ($\eta = 0$). We derive closed-form expressions for the angular momentum change and final orbital separation as functions of $\eta$, mass ratio, and transferred fraction, and show that conservative mass transfer corresponds to the unique choice $\vtrans = \vcom$. While the condition $\dot{L}_{\rm orb} = 0$ for conservative transfer is well established \citep{FrankKingRaine2002, Tauris2023}, and \citet{Soberman1997} gave a comprehensive treatment of angular momentum loss modes, the present work makes three specific contributions: (i)~an explicit velocity-based parameterization of the momentum closure during conservative mass transfer, identifying $\vtrans = \vcom$ (equivalently, $\eta = M_1/\mtot$) as the unique physical requirement for angular momentum conservation; (ii)~closed-form expressions for the orbital separation and angular momentum evolution across the full $\eta$ spectrum, with a proof that the orbital response is strictly monotonic in~$\eta$; and (iii)~a structural reinterpretation of the conservative limit as a specific, mass-ratio--dependent point in prescription space rather than a neutral default. The framework operates at the level of secular orbital-update prescriptions, not at the level of hydrodynamic stream modelling; the $\eta$-family does not introduce new physical forces but makes explicit the choice of momentum closure that is implicit in every unresolved mass-transfer calculation. Thus, the novelty is not a new mass-transfer mechanism, but an explicit closure parameter that makes implicit assumptions testable and calibratable. Numerical simulations confirm the analytical results to six significant figures. The $\eta$-family provides a compact framework for characterizing the angular momentum closure in unresolved binary evolution models, and a natural interface for calibrating these models against resolved hydrodynamic simulations.

The paper is organized as follows. Section~\ref{sec2} defines the binary model and transfer geometry. Section~\ref{sec3} introduces the $\eta$-family and derives the core analytical results. Section~\ref{sec4} describes the numerical implementation. Section~\ref{sec5} presents the simulation results, a continuous parameter-space survey across~$\eta$, and convergence tests. Section~\ref{sec6} discusses implications and limitations, and Section~\ref{sec7} summarizes our findings.

\section{Binary Model and Transfer Geometry}\label{sec2}

We consider a circular binary consisting of a donor of mass~$M_1$ and an accretor of mass~$M_2$, with orbital separation~$a$ and mass ratio $q \equiv M_1/M_2$. All calculations are performed in the center-of-mass (COM) frame, in which the positions and velocities satisfy
\begin{equation}\label{eq:com_constraints}
  M_1\mathbf{r}_1 + M_2\mathbf{r}_2 = \mathbf{0}\,,\qquad M_1\mathbf{v}_1 + M_2\mathbf{v}_2 = \mathbf{0}\,,
\end{equation}
so that $\mathbf{r}_2 = -(M_1/M_2)\,\mathbf{r}_1$ and $\mathbf{v}_2 = -(M_1/M_2)\,\mathbf{v}_1$.

Writing $\rell \equiv \mathbf{r}_1 \times \mathbf{v}_1$ for the specific angular momentum at the donor position (a constant for circular orbits), the total orbital angular momentum is
\begin{equation}\label{eq:L_total}
  L = M_1(\mathbf{r}_1 \times \mathbf{v}_1) + M_2(\mathbf{r}_2 \times \mathbf{v}_2) = \frac{M_1\,\mtot}{M_2}\,\rell\,.
\end{equation}

Mass transfer is modeled as a sequence of impulsive events: at each event a mass element $\delta m \ll M_1$ is instantaneously removed from the donor and added to the accretor. The positions $\mathbf{r}_1$, $\mathbf{r}_2$ are unchanged during the mapping; only the masses and velocities are updated. The velocity assigned to the transferred element defines the \emph{momentum prescription} and determines the post-transfer angular momentum.

\section{The $\eta$-Family of Prescriptions}\label{sec3}

We parameterize the velocity of the transferred mass element as a linear combination of the donor and accretor velocities:
\begin{equation}\label{eq:eta_def}
  \vtrans = \eta\,\mathbf{v}_1 + (1-\eta)\,\mathbf{v}_2\,,
\end{equation}
where $\eta \in [0,1]$ is a dimensionless parameter; the restriction to $[0,1]$ ensures that $\vtrans$ is a convex combination of the component velocities, i.e.\ the effective velocity does not extrapolate beyond the donor and accretor velocities in the centre-of-mass frame. Allowing $\eta \notin [0,1]$ is mathematically possible but is not considered here. The transferred element $\delta m$ carries momentum $\delta m \cdot \vtrans$ away from the donor and deposits it onto the accretor, conserving total linear momentum by construction.

\subsection{General angular momentum change}\label{sec3.1}

After the transfer, the donor and accretor momenta are $M_1\mathbf{v}_1 - \delta m\,\vtrans$ and $M_2\mathbf{v}_2 + \delta m\,\vtrans$, respectively. The angular momentum change is therefore $\dL = \delta m\,(\mathbf{r}_2 - \mathbf{r}_1) \times \vtrans$. Using $\mathbf{r}_2 - \mathbf{r}_1 = -(\mtot/M_2)\,\mathbf{r}_1$ and expanding $\vtrans$ (see Appendix~\ref{app1}) gives
\begin{equation}\label{eq:dL_general}
  \frac{\dL}{L} = \delta m\left[\frac{1-\eta}{M_2} - \frac{\eta}{M_1}\right]\,.
\end{equation}
This is exact for circular orbits with impulsive transfer and is independent of orbital phase.

\subsection{Endpoint prescriptions}\label{sec3.2}

\paragraph{Donor-velocity prescription ($\eta = 1$).}
The transferred mass carries the donor's instantaneous velocity~$\mathbf{v}_1$. The accretor velocity updates as
\begin{equation}\label{eq:v2_update}
  \mathbf{v}_{2,\mathrm{new}} = \frac{M_2\,\mathbf{v}_2 + \delta m\,\mathbf{v}_1}{M_2 + \delta m}\,,
\end{equation}
and Equation~(\ref{eq:dL_general}) reduces to
\begin{equation}\label{eq:dL_donor}
  \frac{\dL_{\rm d}}{L} = -\frac{\delta m}{M_1}\,.
\end{equation}
The angular momentum \emph{always decreases}, regardless of mass ratio.

\paragraph{Accretor-velocity prescription ($\eta = 0$).}
The transferred mass is assigned the accretor's velocity~$\mathbf{v}_2$. The donor velocity updates as
\begin{equation}\label{eq:v1_update}
  \mathbf{v}_{1,\mathrm{new}} = \frac{M_1\,\mathbf{v}_1 - \delta m\,\mathbf{v}_2}{M_1 - \delta m}\,,
\end{equation}
and Equation~(\ref{eq:dL_general}) gives
\begin{equation}\label{eq:dL_accretor}
  \frac{\dL_{\rm a}}{L} = +\frac{\delta m}{M_2}\,.
\end{equation}
The angular momentum \emph{always increases}. The ratio of the two endpoint changes is
\begin{equation}\label{eq:ratio}
  \frac{\dL_{\rm a}}{\dL_{\rm d}} = -\,q\,,
\end{equation}
confirming opposite signs with magnitude ratio equal to the mass ratio.

\subsection{Conservative transfer as a special case}\label{sec3.3}

Setting $\dL = 0$ in Equation~(\ref{eq:dL_general}) and solving for $\eta$ gives
\begin{equation}\label{eq:eta_cons}
  \eta_{\rm cons} = \frac{M_1}{\mtot}\,.
\end{equation}
Substituting into Equation~(\ref{eq:eta_def}):
\begin{equation}\label{eq:v_cons}
  \vtrans\big|_{\dL=0} = \frac{M_1}{\mtot}\,\mathbf{v}_1 + \frac{M_2}{\mtot}\,\mathbf{v}_2 = \vcom\,.
\end{equation}
Conservative mass transfer---in which the orbital angular momentum is exactly preserved---corresponds uniquely to assigning the transferred material the center-of-mass velocity. This is not a trivial condition: $\eta_{\rm cons}$ depends on the instantaneous mass ratio, so maintaining $\dL = 0$ throughout a transfer sequence requires the effective velocity to track $\vcom$ as the masses evolve.

\subsection{Cumulative effects}\label{sec3.4}

For a transfer of total fraction $f = \Delta M / M_{1,0}$ of the initial donor mass, integrating the per-step expressions yields:
\begin{align}
  \text{Donor-velocity:}\quad \frac{\dL}{L_0} &= -f\,, \label{eq:cumul_donor} \\[4pt]
  \text{Accretor-velocity:}\quad \frac{\dL}{L_0} &= f\,q_0\,, \label{eq:cumul_accretor} \\[4pt]
  \text{Conservative:}\quad \frac{\dL}{L_0} &= 0\,, \label{eq:cumul_cons}
\end{align}
where $q_0 = M_{1,0}/M_{2,0}$ is the initial mass ratio.

\subsection{Mapping to orbital separation}\label{sec3.5}

For a circular orbit, $L = \mu\sqrt{G\,\mtot\,a}$ with reduced mass $\mu = M_1 M_2 / \mtot$. Since $\mtot$ is conserved, the final separation ratio is
\begin{equation}\label{eq:a_general}
  \frac{a_f}{a_0} = \left(\frac{L_f}{L_0}\right)^{\!2} \left(\frac{\mu_0}{\mu_f}\right)^{\!2}.
\end{equation}
Substituting the cumulative $L_f/L_0$ for each prescription gives:
\begin{align}
  \text{Donor-vel.:}\quad \frac{a_f}{a_0} &= \frac{1}{(1 + f\,q_0)^2}\,, \label{eq:a_donor} \\[4pt]
  \text{Accretor-vel.:}\quad \frac{a_f}{a_0} &= \frac{1}{(1-f)^2}\,, \label{eq:a_accretor} \\[4pt]
  \text{Conservative:}\quad \frac{a_f}{a_0} &= \frac{1}{\bigl[(1-f)(1+f\,q_0)\bigr]^2}\,. \label{eq:a_cons}
\end{align}
A notable consequence of Equation~(\ref{eq:a_accretor}) is that the accretor-velocity prescription yields a final separation that depends only on the transferred fraction~$f$, not on the mass ratio. This arises because the $q_0$-dependent factors in $L_f/L_0$ and $\mu_0/\mu_f$ cancel exactly.

\subsection{General constant-$\eta$ solution}\label{sec3.6}

The results above treat only the two endpoints and the conservative case. For arbitrary constant~$\eta$, the cumulative angular momentum change follows from integrating Equation~(\ref{eq:dL_general}) in the continuous limit:
\begin{equation}\label{eq:L_integral}
  \ln\frac{L_f}{L_0} = \int_0^{\Delta M}\!\left[\frac{1-\eta}{M_{2,0}+m} - \frac{\eta}{M_{1,0}-m}\right] dm\,,
\end{equation}
where $M_{1,0}-m$ and $M_{2,0}+m$ are the instantaneous donor and accretor masses after transferring a cumulative amount~$m$. For $\eta$ independent of~$m$, both integrals are elementary, giving
\begin{equation}\label{eq:L_general_eta}
  \frac{L_f}{L_0} = \left(\frac{M_{2,f}}{M_{2,0}}\right)^{\!1-\eta} \left(\frac{M_{1,f}}{M_{1,0}}\right)^{\!\eta} = (1 + f\,q_0)^{1-\eta}\,(1-f)^{\eta}\,.
\end{equation}
Combining with $\mu_0/\mu_f = [(1-f)(1+f\,q_0)]^{-1}$ from Equation~(\ref{eq:a_general}) yields the general closed-form separation ratio:
\begin{equation}\label{eq:a_general_eta}
  \frac{a_f}{a_0} = \frac{1}{(1-f)^{\,2(1-\eta)}\,(1+f\,q_0)^{\,2\eta}}\,.
\end{equation}
Setting $\eta = 1$ recovers Equation~(\ref{eq:a_donor}); $\eta = 0$ recovers Equation~(\ref{eq:a_accretor}). Table~\ref{tab1} includes this general result alongside the three special cases.

Note that exact angular momentum conservation requires $\eta = \eta_{\rm cons}(q) = M_1/\mtot$, which varies as mass is transferred. The constant-$\eta$ solution in Equation~(\ref{eq:a_general_eta}) therefore does not reduce to the conservative result (Equation~\ref{eq:a_cons}) for $\eta = M_{1,0}/\mtot$, except in the limit $f \to 0$. The conservative formula is instead recovered by allowing $\eta$ to track $M_1/\mtot$ at each step, which is equivalent to setting $\dL = 0$ exactly throughout the integration.

Table~\ref{tab1} summarizes the key formulae.

\begin{table}[t]
\centering
\caption{Summary of analytical results. $f$~is the fractional donor mass transferred; $q_0 = M_{1,0}/M_{2,0}$ is the initial mass ratio.\label{tab1}}
\begin{tabular*}{\columnwidth}{@{\extracolsep\fill}lccc@{\extracolsep\fill}}
\toprule
Prescription & $\dL/L$ & $L_f/L_0$ & $a_f/a_0$ \\
 & (per step) & (cumul.) & \\
\midrule
Donor-vel. & $-\delta m/M_1$ & $1-f$ & $(1+fq_0)^{-2}$ \\
Accretor-vel. & $+\delta m/M_2$ & $1+fq_0$ & $(1-f)^{-2}$ \\
Conservative & $0$ & $1$ & $[(1-f)(1+fq_0)]^{-2}$ \\[2pt]
General & Eq.~(\ref{eq:dL_general}) & $(1+fq_0)^{1-\eta}$ & $(1-f)^{-2(1-\eta)}$ \\
(const.~$\eta$) & & $(1-f)^{\eta}$ & $(1+fq_0)^{-2\eta}$ \\
\bottomrule
\end{tabular*}
\end{table}

\subsection{Physical interpretation}\label{sec3.7}

The sign difference between the endpoint prescriptions has a geometric origin. In the COM frame, $\mathbf{v}_1$ and $\mathbf{v}_2$ point in opposite directions: $\mathbf{v}_2 = -(M_1/M_2)\,\mathbf{v}_1$.

Under the donor-velocity prescription, mass arrives at the accretor position carrying a velocity directed \emph{against} the local orbital motion. Incorporating this element reduces the orbital angular momentum, equivalent to an external braking torque. Under the accretor-velocity prescription, mass departs the donor with a velocity opposed to the donor's own orbital motion. Removing this retrograde element increases the orbital angular momentum, equivalent to an external driving torque.

Neither sign is the ``natural'' outcome: both are consequences of unresolved physics encoded in the choice of $\eta$. Conservative transfer ($\eta = \eta_{\rm cons}$) is not a neutral baseline but a specific physical claim---that the stream velocity tracks $\vcom$ at every instant. Any deviation from this condition produces a nonzero effective torque whose sign and magnitude are set by $\eta - \eta_{\rm cons}$.

\begin{figure}[t]
\centering
\includegraphics[width=78mm]{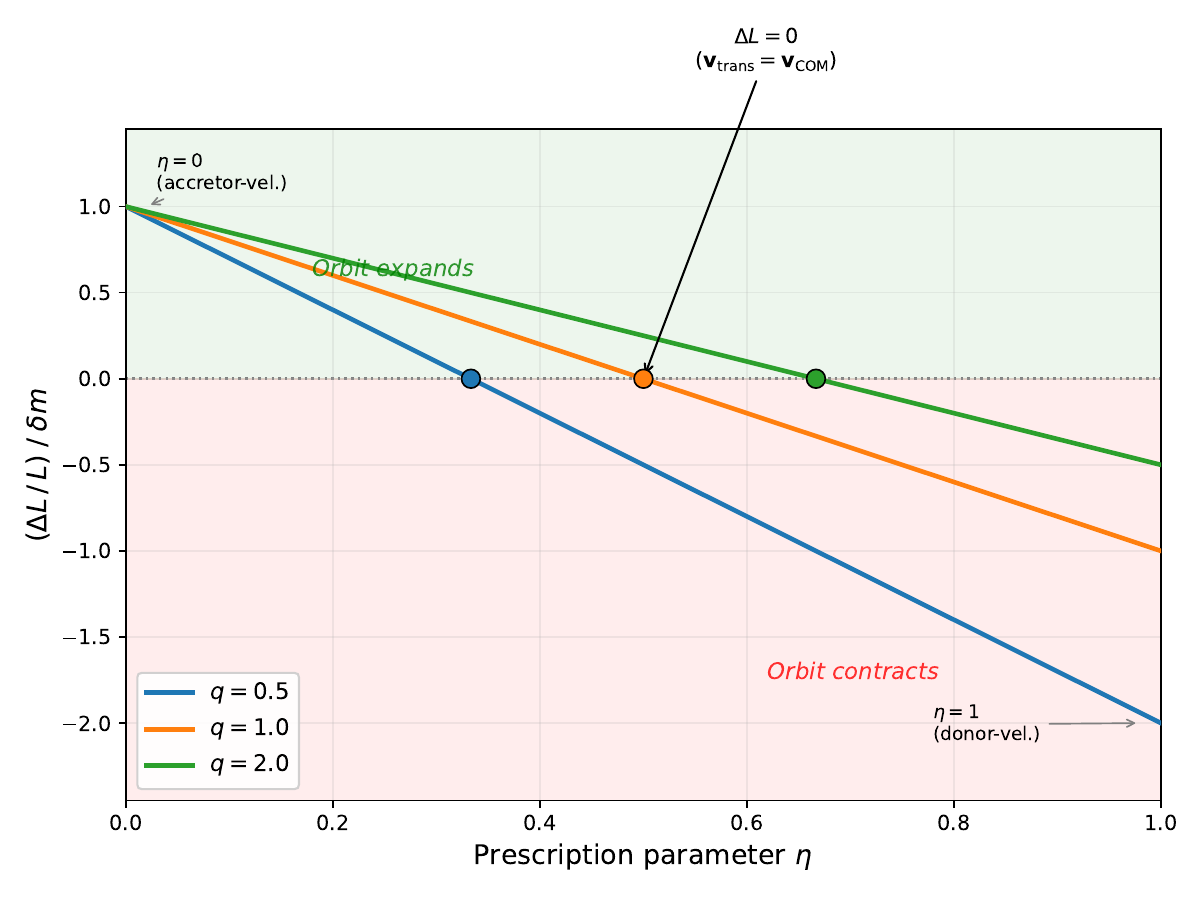}
\caption{Normalized angular momentum change per unit transferred mass as a function of the prescription parameter $\eta$, for three mass ratios (with $M_2 = 1$ normalization, so that $M_1 = q$). Filled circles mark the conservative condition $\eta_{\rm cons} = q/(1+q)$, where $\dL = 0$ and $\vtrans = \vcom$. All three curves converge to the same value at $\eta = 0$, reflecting the fact that the accretor-velocity prescription yields an angular momentum change independent of mass ratio. Shading indicates regions of orbital expansion ($\dL > 0$) and contraction ($\dL < 0$).\label{fig1}}
\end{figure}

Figure~\ref{fig1} illustrates this structure. For each mass ratio $q$, $\dL/L$ is linear in $\eta$, with the zero crossing at $\eta_{\rm cons} = q/(1+q)$ (filled circles). The two endpoint prescriptions always lie on opposite sides of the zero line, producing angular momentum changes of opposite sign.

\section{Numerical Methods}\label{sec4}

We verify the analytical predictions of Section~\ref{sec3} with direct two-body simulations.

\subsection{Integration}\label{sec4.1}

The equations of motion are integrated with the velocity Verlet algorithm, which preserves phase-space volume and ensures bounded energy errors over long integrations. The fiducial resolution is $10^4$~steps per initial orbital period.

\subsection{Transfer implementation}\label{sec4.2}

At each of 10 equally spaced transfer events per orbit, a mass element $\delta m = f\,M_{1,0} / N_{\rm events}$ is instantaneously transferred, with the velocity update given by Equation~(\ref{eq:v2_update}) or~(\ref{eq:v1_update}). After each event the system is recentered to the instantaneous COM frame to suppress floating-point drift in total linear momentum; this does not affect relative orbital diagnostics. The transferred mass is clamped to at most 99.9\% of the current donor mass.

\subsection{Diagnostics}\label{sec4.3}

We record the osculating semi-major axis $a_{\rm osc} = -G\,\mtot/(2\varepsilon)$, where $\varepsilon = v_{\rm rel}^2/2 - G\,\mtot/r$, and the total orbital angular momentum~$L$ at regular intervals. Because transfer is impulsive, $a_{\rm osc}$ changes discontinuously at each event.

\section{Results}\label{sec5}

All figures in this section were generated by the accompanying Python script; parameter choices are listed in the figure captions.

\subsection{Sign reversal at fixed mass ratio}\label{sec5.1}

\begin{figure*}[t]
\centering
\includegraphics[width=342pt]{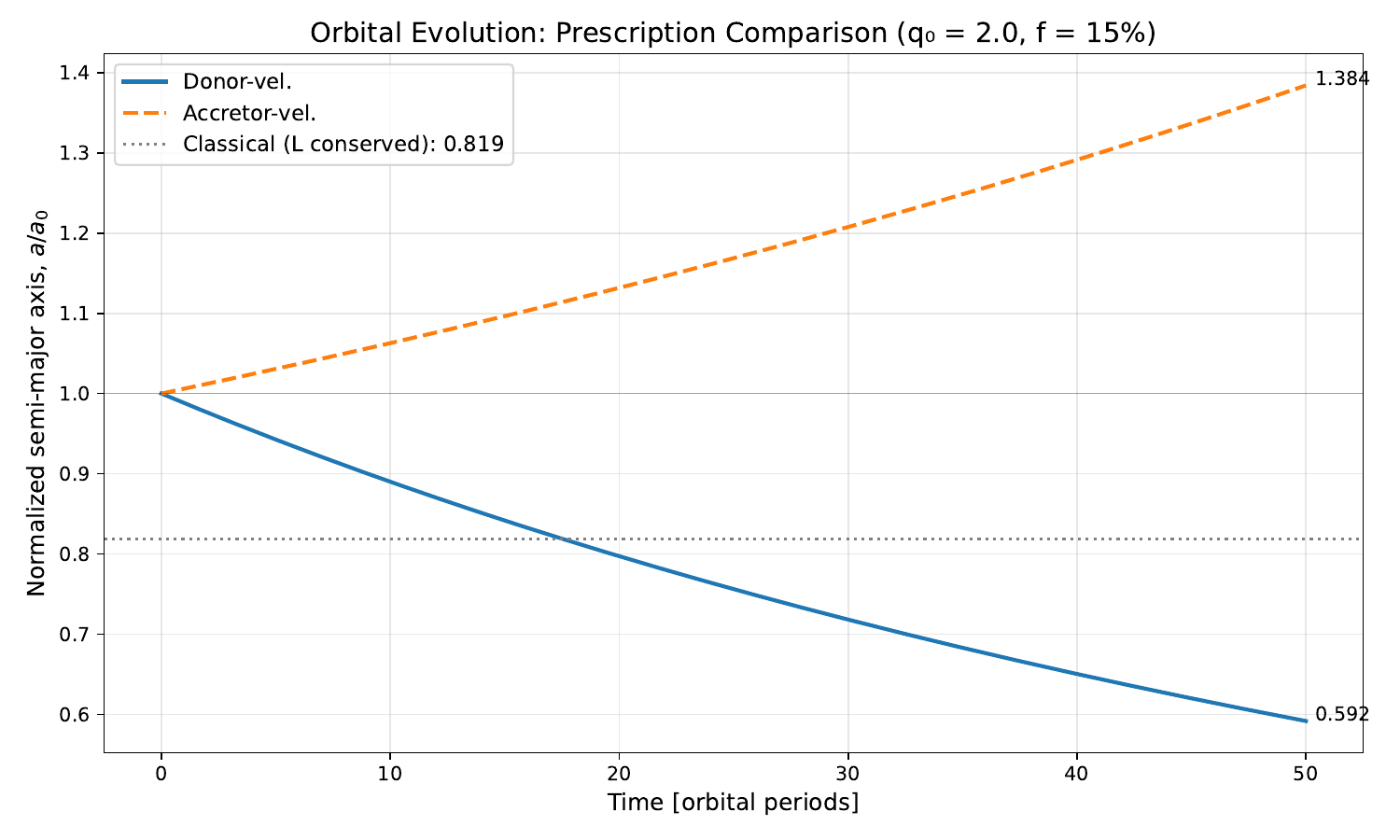}
\caption{Orbital separation vs.\ time for the two endpoint prescriptions at $q_0 = 2.0$, $f = 0.15$, over 50 orbital periods. The donor-velocity prescription ($\eta = 1$; labelled `Donor-vel' in the legend) produces contraction while the accretor-velocity prescription ($\eta = 0$) produces expansion. The classical conservative prediction is shown as a dotted line.\label{fig2}}
\end{figure*}

Figure~\ref{fig2} shows the orbital separation as a function of time for the donor-velocity and accretor-velocity prescriptions at $q_0 = 2.0$ and $f = 0.15$. The two prescriptions produce qualitatively opposite outcomes: the donor-velocity case contracts to $a_f/a_0 = 0.592$, while the accretor-velocity case expands to $a_f/a_0 = 1.384$. The classical (conservative) prediction, $a_f/a_0 = 0.819$, lies between the two.

\begin{figure*}[t]
\centering
\includegraphics[width=342pt]{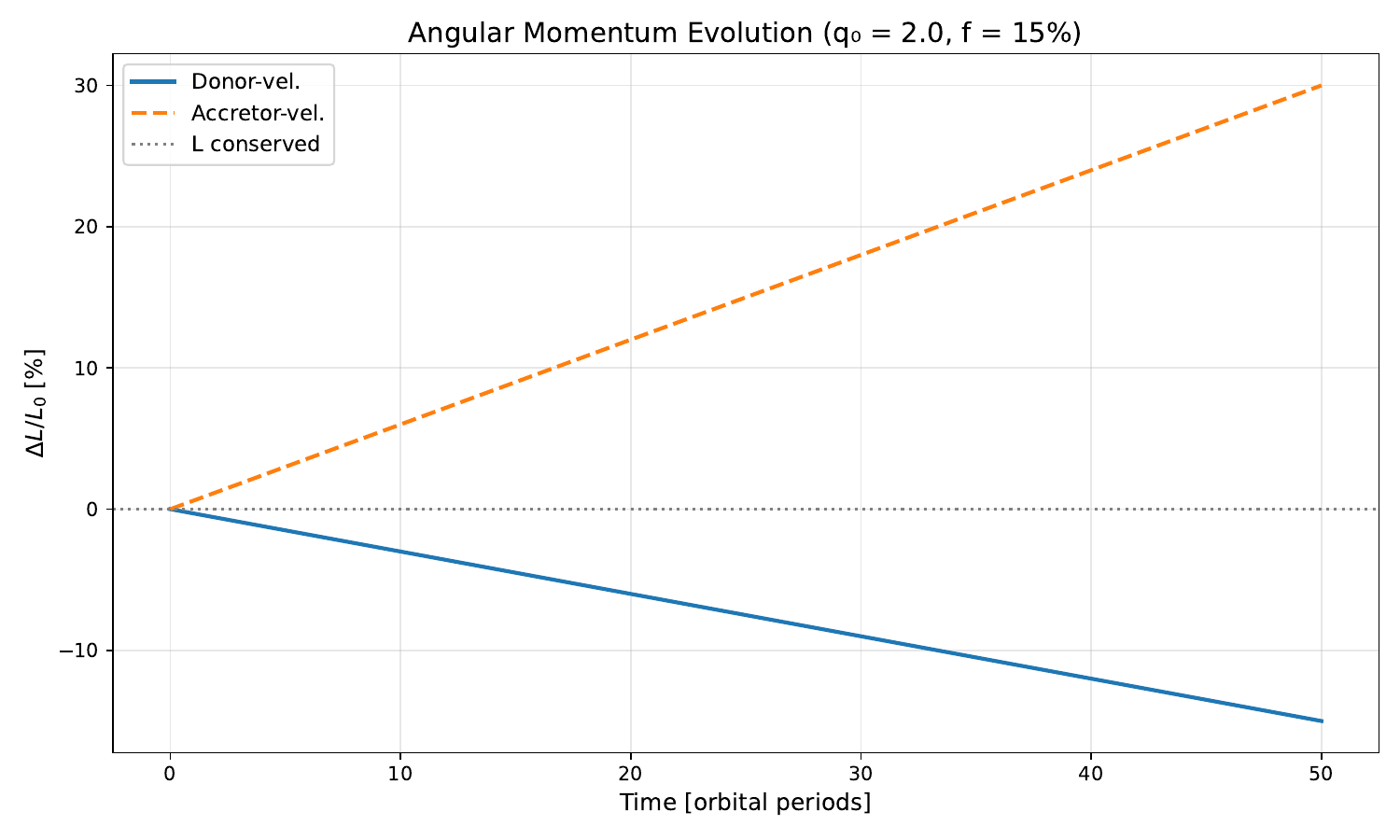}
\caption{Fractional angular momentum change vs.\ time for the same parameters as Figure~\ref{fig2}. The donor-velocity prescription ($\eta = 1$; `Donor-vel.') loses angular momentum while the accretor-velocity prescription ($\eta = 0$) gains it.\label{fig3}}
\end{figure*}

Figure~\ref{fig3} confirms the mechanism: the donor-velocity prescription loses 15\% of the initial angular momentum ($\dL/L_0 = -f$), while the accretor-velocity prescription gains 30\% ($\dL/L_0 = f\,q_0$), in exact agreement with Equations~(\ref{eq:cumul_donor})--(\ref{eq:cumul_accretor}).

\subsection{Parameter-space survey}\label{sec5.2}

\begin{figure*}[t]
\centering
\includegraphics[width=342pt]{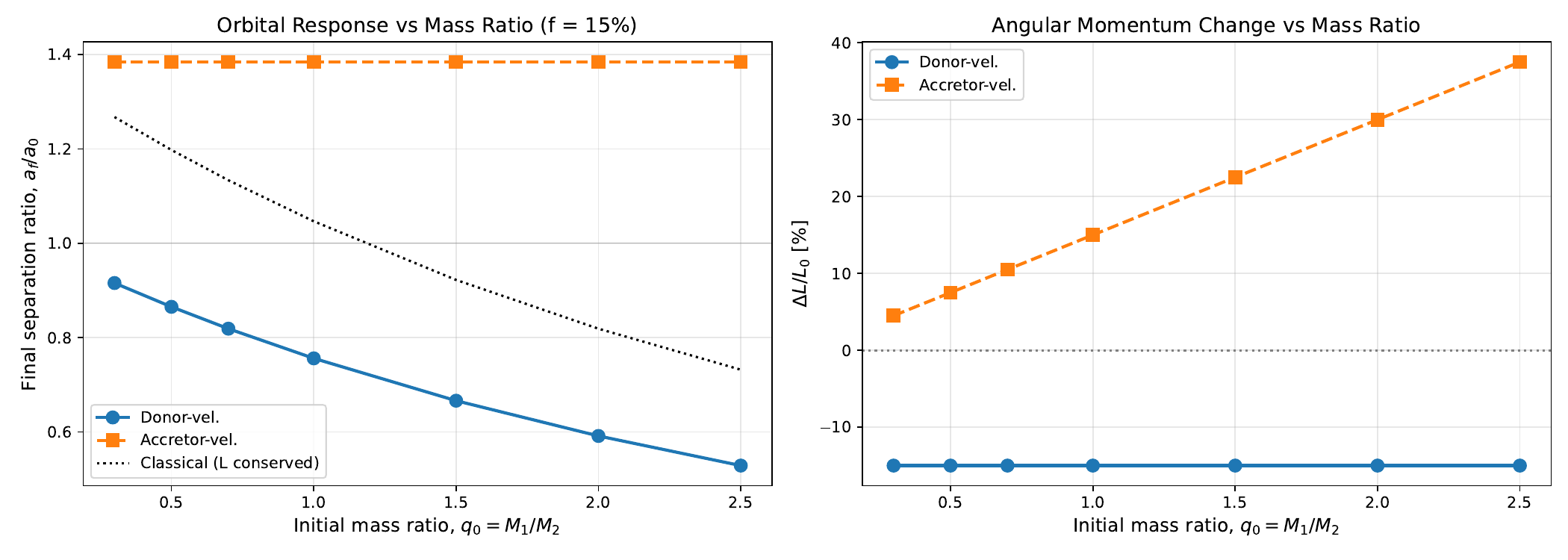}
\caption{Left: final separation ratio $a_f/a_0$ vs.\ initial mass ratio $q_0$ for the two endpoint prescriptions and the conservative case. Right: fractional angular momentum change $\dL/L_0$ vs.\ $q_0$. Transfer fraction $f = 0.15$ in all cases. The sign reversal persists at all mass ratios tested.\label{fig4}}
\end{figure*}

Figure~\ref{fig4} shows $a_f/a_0$ and $\dL/L_0$ as functions of initial mass ratio for both endpoint prescriptions and the conservative case. The donor-velocity curves depend strongly on $q_0$, while the accretor-velocity curves are independent of $q_0$---a direct consequence of Equation~(\ref{eq:a_accretor}). The sign reversal persists across the entire range $q_0 \in [0.3, 2.5]$.

The spread between the two endpoint predictions provides an upper bound on the uncertainty in $a_f/a_0$ due to the momentum closure. At $q_0 = 2$ and $f = 0.15$, this spread is $\Delta(a_f/a_0) = 1.384 - 0.592 = 0.79$, representing an 80\% range in the final orbital separation.

\subsection{Continuous dependence on $\eta$}\label{sec5.3}

\begin{figure*}[t]
\centering
\includegraphics[width=342pt]{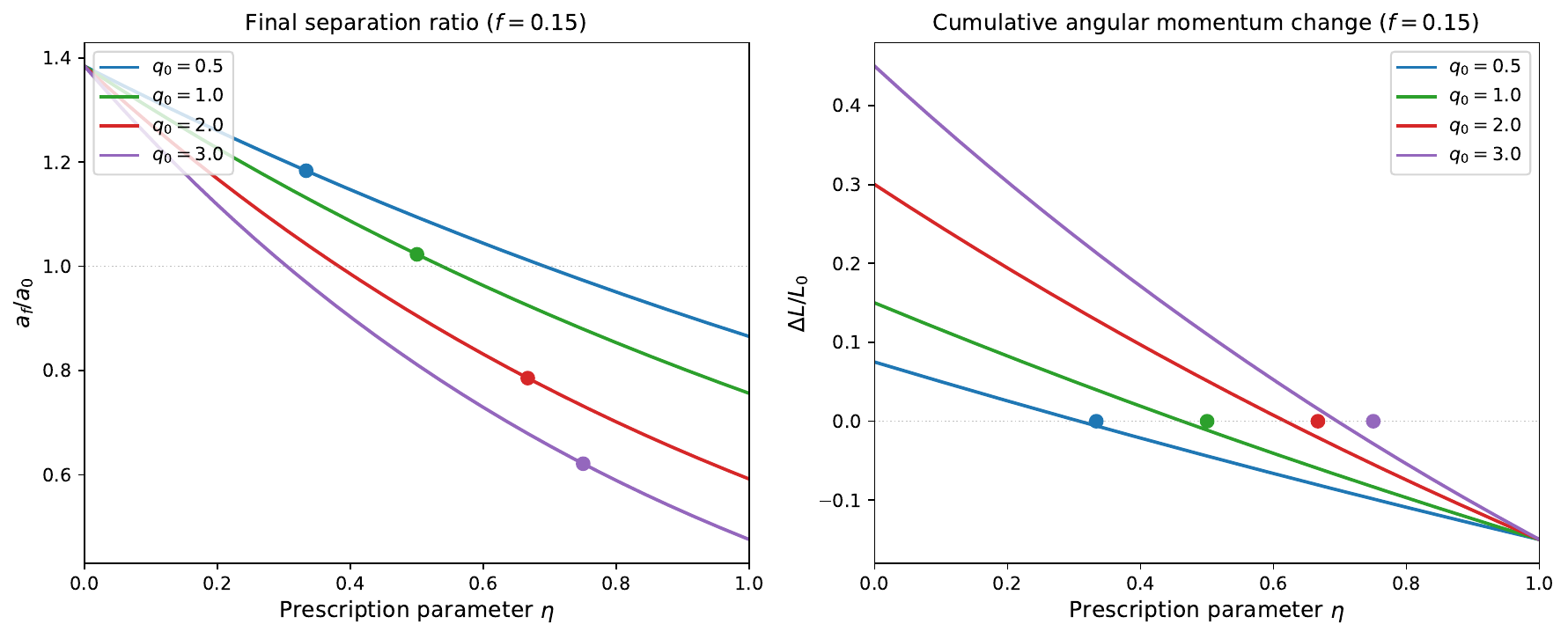}
\caption{Continuous dependence of orbital evolution on the prescription parameter~$\eta$, evaluated from the general closed-form solution (Equation~\ref{eq:a_general_eta}) at $f = 0.15$. Left: final separation ratio $a_f/a_0$ vs.\ $\eta$ for four initial mass ratios. At $\eta = 0$ all curves converge because the accretor-velocity prescription yields $a_f/a_0 = (1-f)^{-2}$, independent of~$q_0$ (Equation~\ref{eq:a_accretor}). Right: cumulative fractional angular momentum change $\dL/L_0$ vs.\ $\eta$, evaluated analytically from Equation~(\ref{eq:L_general_eta}). Filled circles mark the conservative condition $\eta_{\rm cons} = q_0/(1+q_0)$ for each mass ratio. For fixed $f$ and $q_0$, the final separation is a strictly monotonic function of~$\eta$; the two endpoint prescriptions ($\eta = 0, 1$) are the extremes of a continuous family.\label{fig5}}
\end{figure*}

The preceding figures compare only the two endpoint prescriptions and the conservative case. To demonstrate that these are not isolated special cases but part of a continuous spectrum, Figure~\ref{fig5} evaluates the general constant-$\eta$ solution (Equation~\ref{eq:a_general_eta}) across the full range $\eta \in [0,1]$ for several initial mass ratios.

The left panel shows the final separation ratio $a_f/a_0$ as a function of~$\eta$ at fixed $f = 0.15$, for $q_0 = 0.5$, 1.0, 2.0, and~3.0. The two endpoints ($\eta = 0$ and $\eta = 1$) correspond to the accretor-velocity and donor-velocity prescriptions, respectively, and the conservative result is recovered at the mass-ratio--dependent point $\eta_{\rm cons} = q_0/(1+q_0)$ (marked by filled circles). The right panel shows the corresponding cumulative angular momentum change $\dL/L_0$. The zero crossing in each curve coincides with the conservative point. For $\eta > \eta_{\rm cons}$ the orbit contracts; for $\eta < \eta_{\rm cons}$ it expands. The curves confirm that the endpoint prescriptions bracket all intermediate behaviours, and that the conservative case is not a default but a specific point on a continuous axis. For fixed $f$ and $q_0$, the final separation is a strictly monotonic function of~$\eta$: the $\eta$-family makes explicit that conservative transfer is not a physical inevitability but a momentum-closure choice, and any departure from $\eta_{\rm cons}$ produces a monotonic shift in the orbital outcome. This monotonicity is not particular to the value $f = 0.15$ used in the figure. From Equation~(\ref{eq:a_general_eta}), $\partial\ln(a_f/a_0)/\partial\eta = 2\ln(1-f) - 2\ln(1+fq_0)$, which is strictly negative for all $f \in (0,1)$ and $q_0 > 0$ because $(1-f) < 1$ and $(1+fq_0) > 1$. The monotonic structure therefore holds for arbitrary transfer fractions and mass ratios within the assumptions of Section~\ref{sec6.4}.

\subsection{Convergence}\label{sec5.4}

Table~\ref{tab2} demonstrates that the results are insensitive to both the transfer frequency and the integration resolution. $a_f/a_0$ and $\dL/L_0$ are stable to six significant figures across all parameter combinations, confirming that the sign reversal is determined by the analytical prescription, not by discretization.

\begin{table}[t]
\centering
\caption{Convergence test for $q_0 = 2.0$, $f = 0.15$, 50 orbits.\label{tab2}}
\begin{tabular*}{\columnwidth}{@{\extracolsep\fill}rcccc@{\extracolsep\fill}}
\toprule
& \multicolumn{2}{c}{Donor-vel.} & \multicolumn{2}{c}{Accretor-vel.} \\
\cmidrule(lr){2-3}\cmidrule(lr){4-5}
Param. & $a_f/a_0$ & $\dL/L_0$ & $a_f/a_0$ & $\dL/L_0$ \\
\midrule
\multicolumn{5}{l}{\textit{Transfers/orbit (steps $= 10^4$)}} \\
5 & 0.591720 & $-$15.00\% & 1.384087 & $+$30.00\% \\
10 & 0.591719 & $-$15.00\% & 1.384087 & $+$30.00\% \\
20 & 0.591719 & $-$15.00\% & 1.384087 & $+$30.00\% \\
50 & 0.591719 & $-$15.00\% & 1.384087 & $+$30.00\% \\
\midrule
\multicolumn{5}{l}{\textit{Steps/orbit (transfers $= 10$)}} \\
2000 & 0.591719 & $-$15.00\% & 1.384087 & $+$30.00\% \\
5000 & 0.591719 & $-$15.00\% & 1.384087 & $+$30.00\% \\
10000 & 0.591719 & $-$15.00\% & 1.384087 & $+$30.00\% \\
20000 & 0.591719 & $-$15.00\% & 1.384087 & $+$30.00\% \\
\bottomrule
\end{tabular*}
\end{table}

\section{Discussion}\label{sec6}

\subsection{Implications for population synthesis}\label{sec6.1}

The condition for conservative mass transfer---that the transferred mass carry the centre-of-mass velocity---is implicit in the standard derivation of Equation~(\ref{eq:classical}) \citep[e.g.][Ch.~4]{FrankKingRaine2002, Tauris2023}. However, this condition is typically expressed in terms of orbital angular momentum conservation ($\dot{L}_{\rm orb} = 0$), and the underlying velocity requirement is not stated explicitly. The $\eta$-family reformulates this condition as $\eta_{\rm cons} = M_1/\mtot$ and, more importantly, embeds it as a special point within a continuous parameter space. This provides two practical advantages: first, it gives a transparent physical interpretation of what `conservative transfer' means at the level of individual transfer events; and second, it supplies a natural axis along which to quantify the sensitivity of population synthesis predictions to deviations from the conservative assumption.

The $\eta$-family quantifies a previously uncharacterized degree of freedom in binary population synthesis. The spread in $a_f/a_0$ between the endpoint prescriptions (up to 80\% at $q_0 = 2$, $f = 0.15$, depending on the adopted prescription and mass ratio) can be comparable to other leading sources of uncertainty in population synthesis, including common-envelope efficiency and natal kick distributions. Implementations that assume conservative transfer without explicit justification are implicitly setting $\eta = M_1/\mtot$, a mass-ratio-dependent condition that represents a specific physical assumption about the stream dynamics.

In practice, many population synthesis implementations---including BSE \citep{Hurley2002}, COSMIC \citep{Breivik2020}, and COMPAS \citep{TeamCOMPAS2022}---adopt conservative mass transfer via the condition $\dot{L}_{\rm orb} = 0$, which is equivalent to assuming $\vtrans = \vcom$ (i.e., $\eta = M_1/\mtot$) at each update step, whether stated in velocity form or angular-momentum form. This assumption is typically embedded in the orbital update equations rather than exposed as an adjustable parameter. The $\eta$-family makes this implicit assumption explicit and provides a natural axis along which to explore systematic uncertainties: by varying $\eta$ away from $\eta_{\rm cons}$, one can bracket the range of orbital outcomes consistent with different physical assumptions about the stream momentum.

This complements recent work on mass transfer stability criteria \citep{Ge2010, Ge2015, Ge2020}, which characterizes the donor's adiabatic response to rapid mass loss. Together, the donor response (determining \emph{whether} mass transfer is stable) and the momentum prescription (determining \emph{how} the orbit responds to stable transfer) form the two essential ingredients for predicting binary evolution outcomes.

Because the sign of orbital response determines whether a system evolves toward or away from Roche-lobe contact, the prescription choice directly affects predicted rates of stellar mergers, Type Ia supernovae, and compact-object binaries detectable by gravitational-wave observatories \citep{Marchant2016, Mandel2022}. Given that more than 70\% of massive stars interact with a companion during their lifetime \citep{Sana2012}, and that binary interaction dominates the evolution of massive stellar populations \citep{Langer2012}, a systematic survey of population synthesis outcomes across the $\eta$ parameter is needed to bracket this source of uncertainty.

\subsection{Connection to resolved models}\label{sec6.2}

The $\eta$-family provides a natural interface between unresolved prescriptions and resolved hydrodynamic simulations. Given a simulation that tracks the mass-transfer stream through the $L_1$ point \citep{CehulaPejcha2023}, one can compute the effective velocity of accreted material and map it to an effective $\eta$ via Equation~(\ref{eq:eta_def}). This calibrated $\eta$ can then be used directly in population synthesis codes, providing a physically grounded closure without the computational cost of resolving the stream in every system.

A quantitative estimate of the physically expected $\eta$ can be obtained from the hydrodynamics of the mass-transfer stream near~$L_1$, assuming synchronous rotation so that the flow near~$L_1$ corotates with the orbit. The classical result of \citet{LubowShu1975}, confirmed by the nozzle model of \citet{CehulaPejcha2023}, establishes that the stream passes through the $L_1$ point at the local sound speed~$c_s$. Because $c_s \ll v_{\rm orb}$, the stream velocity is dominated by the corotation velocity of~$L_1$ itself; the radial (along the line of centres) component~$c_s$ contributes negligibly to the orbital angular momentum budget. The tangential velocity of material at~$L_1$ is therefore, to leading order, $v_{L_1} \approx \omega\,x_{L_1}$, where $x_{L_1}$ is the distance from the centre of mass to~$L_1$.

Writing $x_{L_1} \approx x_1 - R_L$, where $x_1 = M_2\,a/\mtot$ is the donor's distance from the centre of mass and $R_L$ is the effective Roche-lobe radius \citep{Eggleton1983}---here used as an order-of-magnitude proxy for the distance from the donor centre to~$L_1$---we map the corotation velocity to~$\eta$ as follows. The centre-of-mass constraint $\mathbf{v}_2 = -(M_1/M_2)\,\mathbf{v}_1$ gives the tangential component of Equation~(\ref{eq:eta_def}) as $v_{\rm trans} = v_1[\eta - (1-\eta)\,q]$. Setting $v_{\rm trans} = \omega\,x_{L_1}$ and $v_1 = \omega\,x_1$ yields $\eta(1+q) = q + x_{L_1}/x_1$. Substituting $x_{L_1}/x_1 = 1 - R_L/x_1 = 1 - (R_L/a)(1+q)$, the $(1+q)$ factors cancel exactly, giving
\begin{equation}\label{eq:eta_eff}
  \eta_{\rm eff}(q) \approx 1 - \frac{R_L(q)}{a}\,,
\end{equation}
where $R_L/a$ is given by the \citet{Eggleton1983} fitting formula
\begin{equation}\label{eq:eggleton}
  \frac{R_L}{a} = \frac{0.49\,q^{2/3}}{0.6\,q^{2/3} + \ln(1 + q^{1/3})}\,, \qquad q \equiv \frac{M_1}{M_2}\,.
\end{equation}

\begin{figure}
\centering
\includegraphics[width=\columnwidth]{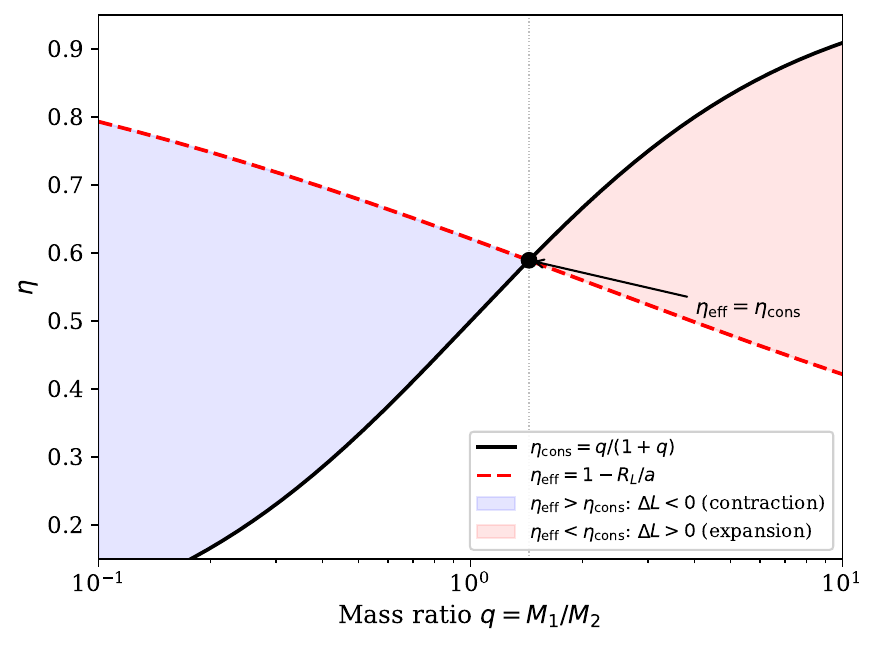}
\caption{Corotation estimate $\eta_{\rm eff}(q) = 1 - R_L/a$ (dashed) compared with the conservative condition $\eta_{\rm cons}(q) = q/(1+q)$ (solid) as functions of mass ratio. The filled circle marks their intersection at $q$ of order unity. Blue shading indicates the region where $\eta_{\rm eff} > \eta_{\rm cons}$, corresponding to $\dL < 0$ (orbital contraction); red shading indicates $\eta_{\rm eff} < \eta_{\rm cons}$, corresponding to $\dL > 0$ (orbital expansion).\label{fig:eta_comparison}}
\end{figure}

Table~\ref{tab3} and Figure~\ref{fig:eta_comparison} compare $\eta_{\rm eff}$ with the conservative value $\eta_{\rm cons} = q/(1+q)$. The corotation estimate yields $\eta_{\rm eff} \approx 0.5$--$0.7$ for common interacting-binary mass ratios ($q \sim 0.3$--$3$), and may lie on either side of $\eta_{\rm cons}$ depending on~$q$: within the $\eta$-parameterization, $\eta_{\rm eff} > \eta_{\rm cons}$ corresponds to $\dL < 0$ and orbital contraction, while $\eta_{\rm eff} < \eta_{\rm cons}$ corresponds to $\dL > 0$ and orbital expansion. The two functions intersect at a mass ratio of order unity; only in the vicinity of this intersection does the corotation limit approximately coincide with the conservative condition. This demonstrates that even the idealised corotation limit at~$L_1$ does not generically correspond to conservative transfer, and that the sign of the orbital torque depends on the mass ratio through the Roche-lobe geometry.

\begin{table}
\caption{Corotation estimate $\eta_{\rm eff} \approx 1 - R_L/a$ compared with the conservative value $\eta_{\rm cons} = q/(1+q)$ for representative mass ratios.\label{tab3}}
\begin{tabular*}{\columnwidth}{@{\extracolsep\fill}cccc@{\extracolsep\fill}}
\toprule
$q = M_1/M_2$ & $R_L/a$ & $\eta_{\rm eff}$ & $\eta_{\rm cons}$ \\
\midrule
0.3 & 0.253 & 0.747 & 0.231 \\
0.5 & 0.326 & 0.674 & 0.333 \\
1.0 & 0.378 & 0.622 & 0.500 \\
1.5 & 0.414 & 0.586 & 0.600 \\
2.0 & 0.440 & 0.560 & 0.667 \\
3.0 & 0.476 & 0.524 & 0.750 \\
\bottomrule
\end{tabular*}
\end{table}

This estimate should be regarded as a leading-order heuristic: it assumes synchronous rotation of the donor, neglects the thermal velocity component ($\sim c_s$), and neglects the deflection of the stream by the Coriolis force after leaving~$L_1$ \citep[see][for detailed ballistic trajectory calculations]{LubowShu1975}. In disk-mediated systems, the effective~$\eta$ may be further modified by circularisation and viscous angular momentum transport. In principle, $\eta$ can be calibrated against hydrodynamic simulations by measuring the mass-weighted stream momentum at impact or accretion and mapping the result to Equation~(\ref{eq:eta_def}). Nonetheless, Equation~(\ref{eq:eta_eff}) provides a physically motivated order-of-magnitude reference value that can serve as a starting point for such calibrations. We emphasize that this estimate is illustrative; it is not a substitute for calibration against resolved simulations of specific systems.

The accretor-velocity endpoint ($\eta = 0$) is the least physically motivated but serves as a mathematical bound that makes the family complete; this limit is included for completeness and to delineate the bounds of the framework.

\subsection{Relation to non-conservative formalisms}\label{sec6.3}

The Soberman et al.\ formalism and the $\eta$-family operate on complementary aspects of the angular momentum budget. In the notation of \citet{Soberman1997}, their parameter $\alpha$ represents isotropic re-emission from the accretor (mass lost carrying the specific angular momentum of the accretor), $\beta$ represents Jeans-mode ejection (mass lost carrying the specific angular momentum of the donor), and $\delta$ represents mass lost to a circumbinary ring. These modes describe angular momentum carried \emph{out of} the system by non-accreted mass. Our $\eta$, by contrast, describes the momentum partition of mass that \emph{remains} in the system and is transferred from donor to accretor.

In a general non-conservative scenario, a fraction $(1 - \beta_{\rm acc})$ of the transferred mass is lost from the system (where $\beta_{\rm acc}$ is the mass transfer efficiency as defined by \citealt{deMink2009}), and the Soberman parameters govern its angular momentum content. For the remaining fraction $\beta_{\rm acc}$ that is accreted, $\eta$ governs the momentum partition. A complete treatment would therefore require specifying both: the Soberman parameters for the lost mass, and $\eta$ for the retained mass. The conservative limit corresponds to $\beta_{\rm acc} = 1$ and $\eta = M_1/\mtot$.

\subsection{Limitations and scope of applicability}\label{sec6.4}

The analytical framework developed here rests on several simplifying assumptions, which we summarize explicitly:
\begin{itemize}
  \item \textbf{Circular orbits.} The per-step $\dL/L$ (Equation~\ref{eq:dL_general}) is exact for circular orbits, where the specific angular momentum $\rell$ is constant and independent of orbital phase. For eccentric orbits, $\rell$ varies with true anomaly and the cumulative result depends on the phasing of transfer events; an extension is straightforward numerically but requires care analytically.
  \item \textbf{Impulsive, instantaneous transfer.} Mass is relocated between point-mass components in discrete steps with no transit time. In reality the mass-transfer stream has finite extent and travel time, and may interact with an accretion disk before being incorporated into the accretor.
  \item \textbf{No spin angular momentum exchange.} We track only the orbital angular momentum. Tidal coupling and spin-orbit interaction can transfer angular momentum between the orbit and stellar spins, modifying the effective $\eta$ on secular timescales.
  \item \textbf{No systemic mass loss.} All transferred mass is retained by the accretor ($\beta_{\rm acc}=1$). Non-conservative transfer with mass loss from the system (e.g.\ through $L_2$ overflow) introduces additional angular momentum loss channels \citep[parameterized by the $\alpha$, $\beta$, $\delta$ modes of][]{Soberman1997} and requires the combined treatment outlined in Section~\ref{sec6.3}.
  \item \textbf{No gravitational torque from the extended stream.} The stream is treated as having zero spatial extent; in reality its gravitational back-reaction on the orbit is nonzero and may be non-negligible in some regimes.
\end{itemize}
A systematic exploration of the joint ($\eta$, $\beta_{\rm acc}$) parameter space, relaxing the conservative assumption, is deferred to future work, especially for eccentric binaries and disk-mediated accretion systems.

\section{Summary}\label{sec7}

We have introduced a one-parameter family of momentum prescriptions for mass transfer in circular binaries. The parameter $\eta$ specifies the effective velocity of the transferred mass as a linear combination of the donor and accretor velocities. We denote this parameter $\eta$ to distinguish it from the mass-loss mode parameters ($\alpha$, $\beta$, $\delta$) of \citet{Soberman1997}, which govern non-conservative angular momentum losses.

Our principal results are:
\begin{enumerate}[1.]
  \item The per-step angular momentum change is $\dL/L = \delta m\,[(1-\eta)/M_2 - \eta/M_1]$, exact for circular orbits with impulsive transfer (Equation~\ref{eq:dL_general}).
  \item The two endpoint prescriptions ($\eta = 1$ and $\eta = 0$) produce angular momentum changes of opposite sign at every mass ratio, leading to contraction versus expansion of the orbit (Equations~\ref{eq:dL_donor}--\ref{eq:dL_accretor}).
  \item Conservative mass transfer ($\dL = 0$) is recovered uniquely when the transferred mass carries the center-of-mass velocity, $\eta = M_1/\mtot$ (Equation~\ref{eq:eta_cons}).
  \item For constant~$\eta$, the general closed-form solution is $a_f/a_0 = {(1-f)^{-2(1-\eta)}}\,{(1+fq_0)^{-2\eta}}$, which recovers all endpoint cases (Equation~\ref{eq:a_general_eta}).
  \item A corotation estimate based on the tangential velocity of the $L_1$ point gives $\eta_{\rm eff} \approx 1 - R_L/a$, which yields $\eta_{\rm eff} \approx 0.5$--$0.7$ for typical mass ratios and may lie on either side of~$\eta_{\rm cons}$ depending on~$q$ (Equation~\ref{eq:eta_eff}; Table~\ref{tab3}).
\end{enumerate}

These results provide a unified analytical framework for quantifying the sensitivity of binary orbital evolution to the momentum closure, and a bridge between unresolved population synthesis prescriptions and resolved hydrodynamic simulations. The simulation code is available at \url{https://github.com/fireheartjerry/eta-family-mass-transfer}.

\section*{Acknowledgments}

The author thanks Alexander Mushtukov for guidance and supervision throughout this research, Hongwei Ge for valuable discussions on mass transfer stability criteria, and Jakub Cehula for helpful comments on the manuscript. AI-assisted tools were used for language editing.

\subsection*{Conflict of interest}

The author declares no potential conflict of interests.

\subsection*{Data availability statement}

The simulation code and data underlying this article are publicly available at \url{https://github.com/fireheartjerry/eta-family-mass-transfer}.

\appendix

\section{Derivation of the General $\dL$ Formula}\label{app1}

We derive Equation~(\ref{eq:dL_general}) from first principles.

Before transfer, the total angular momentum is (Equation~\ref{eq:L_total}):
\begin{equation}
  L = \frac{M_1\,\mtot}{M_2}\,\rell\,.
\end{equation}

After transfer, the donor has momentum $M_1\mathbf{v}_1 - \delta m\,\vtrans$ and the accretor has $M_2\mathbf{v}_2 + \delta m\,\vtrans$. The post-transfer angular momentum is
\begin{align}
  L_{\rm after} &= \mathbf{r}_1 \times \bigl[M_1\mathbf{v}_1 - \delta m\,\vtrans\bigr] + \mathbf{r}_2 \times \bigl[M_2\mathbf{v}_2 + \delta m\,\vtrans\bigr] \notag\\
    &= L + \delta m\,(\mathbf{r}_2 - \mathbf{r}_1) \times \vtrans\,.
\end{align}

Using $\mathbf{r}_2 - \mathbf{r}_1 = -(\mtot/M_2)\,\mathbf{r}_1$ and expanding $\vtrans = \eta\,\mathbf{v}_1 + (1-\eta)\,\mathbf{v}_2$:
\begin{align}
  \dL &= -\frac{\delta m\,\mtot}{M_2} \bigl[\eta\,\rell - (1-\eta)\,\tfrac{M_1}{M_2}\,\rell\bigr] \notag\\
      &= -\frac{\delta m\,\mtot\,\rell}{M_2^2} \bigl[\eta\,M_2 - (1-\eta)\,M_1\bigr]\,.
\end{align}

Dividing by $L = M_1\,\mtot\,\rell / M_2$:
\begin{equation}
  \frac{\dL}{L} = -\frac{\delta m}{M_1\,M_2} \bigl[\eta\,M_2 - (1-\eta)\,M_1\bigr] = \delta m\!\left[\frac{1-\eta}{M_2} - \frac{\eta}{M_1}\right].
\end{equation}

Setting $\eta = 1$ gives $-\delta m/M_1$; $\eta = 0$ gives $+\delta m/M_2$; $\dL = 0$ gives $\eta = M_1/\mtot$. \hfill$\square$

\bibliography{refs}

\end{document}